%% file: main.tex
\documentclass{aa}  

\usepackage{graphicx}
\usepackage{txfonts}

\DeclareGraphicsExtensions{.pdf,.png}

\begin{document}

   \title{Frequencies, chaos and resonances: a study of orbital parameters of nearby thick disc and halo stars}
   \titlerunning{Chaos and resonances}
   \author{Helmer H. Koppelman
          \and
          Jorrit H. J. Hagen
          \and
          Amina Helmi
          }

   \institute{Kapteyn Astronomical Institute, University of Groningen, Landleven 12, 9747 AD Groningen, The Netherlands\\
              \email{koppelman@astro.rug.nl}
}
   \date{}

  \abstract
   {}
   {
    We study the distribution of nearby thick disc and halo stars in subspaces defined by their characteristic orbital parameters. Our aim is to establish the origin of the structure reported in particular, in the $R_{\rm max}-z_{\rm max}$ space.
   }
   {
   To this end, we compute the orbital parameters and frequencies of stars for a generic and for a St\"ackel Milky Way potential.
   }
   {
   We find that for both the thick disc and halo populations very similar prominent substructures are apparent for the generic Galactic potential, while no substructure is seen for the St\"ackel model. This indicates that the origin of these features is not merger-related, but due to non-integrability of the generic potential. This conclusion is strengthened by our frequency analysis of the orbits of stars, which reveals the presence of prominent resonances, with $\sim 30\%$ of the halo stars associated to resonance families. In fact, the stars in resonances define the substructures seen in the spaces of characteristic orbital parameters. Furthermore, we find that some stars in our sample and in debris streams are on the same resonance as the Sagittarius dwarf, suggesting this system has also influenced the distribution of stars in the Galactic thick disc and halo components.
   }
   {
   Our study constitutes a step towards disentangling the imprint of merger debris from substructures driven by internal dynamics. Given their prominence, these resonant-driven overdensities could potentially be useful to constrain the exact form of the Galactic potential.
   }


   \maketitle
%

\input{Introduction}
\input{Data}

\input{Results}
\input{Analysis}
\input{Discussion}

\input{Conclusions}

\begin{acknowledgements}
We gratefully acknowledge financial support from a VICI grant from the Netherlands Organisation for Scientific Research (NWO) and a Spinoza prize. This work has made use of data from the European Space Agency (ESA) mission Gaia (\url{http://www.cosmos.esa.int/gaia}), processed by the Gaia Data Processing and Analysis Consortium (DPAC, \url{http://www.cosmos.esa.int/web/gaia/dpac/consortium}). Funding for the DPAC has been provided by national institutions, in particular the institutions participating in the Gaia Multilateral Agreement. 
For the analysis, the following software packages have been used: {\tt vaex} \citep{Breddels2018}, {\tt numpy} \citep{VanDerWalt2011TheComputation}, {\tt matplotlib} \citep{Hunter2007Matplotlib:Environment}, {\tt jupyter notebooks} \citep{Kluyver2016JupyterWorkflows}.
\end{acknowledgements}

\bibliographystyle{aa} 
\bibliography{main.bib} 


\end{document}

%% file: Introduction.tex
\section{Introduction}\label{sec:intro}
With the second data release of the {\it Gaia} space mission \citep{GaiaDR2_summary_Brownetal2018}, a catalogue comprising the full six-dimensional (6D) phase-space
information has become available \citep{Katz2019GaiaVelocities}. This subset has enabled detailed studies of interconnected dynamical processes taking place in the Milky Way. For example, substructures in integrals-of-motion spaces of nearby disc stars, have been related to the Galactic bar or spiral arms \citep[e.g.][]{Dehnen2000, Monarietal2017, Khoperskovetal2019, Trick2019TheDR2, Hunt2019SignaturesDisc, Monari2019SignaturesSpace}, or to interactions with satellite galaxies \citep{Antojaetal2018, Laporteetal2018_Sgr&LMC, Laporteetal2019}.

Also the assembly history of the Milky Way can be studied by identifying and characterising substructures in integrals-of-motion space, as first put forward by \cite{Helmi&deZweeuw2000}. Structures comprising tens to hundreds of stars have been linked to accreted dwarf galaxies whose debris can still be traced in the solar neighbourhood \citep[][see also the reviews of \citealt{Newberg2016, Klement2010}]{Helmietal1999_HelmiStreams, Chiba2000, Klement2008, Klement2009a, Williams2011, Helmietal2017, Myeong2018c, Koppelmanetal2018_blob, Koppelmanetal2019, Yuan2019DynamicalHalo, Naidu2020}. Besides these small groups, recently the relics of a massive dwarf galaxy (now known as Gaia-Enceladus) have been identified in the same kind of spaces by \citet[][see also \citealt{Belokurovetal2018}]{Helmietal2018}.

With the rise of machine learning, it has become popular to deploy automated classification algorithms to search for substructures \citep{Necib2019EvidenceVicinity, Borsato2019IdentifyingTechniques, Du2019IdentifyingLearning, Koppelmanetal2019, Ratzenbock2020ExtendedNeighborhood, Yuan2019DynamicalHalo}. Substructures are not necessarily always accreted \citep[e.g.][]{Gomez&Helmi2010, Gomezetal2013_wavesbySgr, Jean-Baptiste2017OnTale} and therefore the interpretation of their origin is not always straightforward. In this work, we investigate structures reported in $R_\mathrm{max} - z_\mathrm{max}$ space \citep[e.g.][]{Haywood2018InDR2} supplemented by information obtained from the orbital frequencies of the stars. 

Substructure identified in the space characterised by $R_\mathrm{max} - z_\mathrm{max}$ may be amenable to intuitive interpretation. \cite{Schuster2012TwoParameters} noted that the low-[$\alpha$/Fe] stars in the halo, identified in \cite{Nissen2010}, have orbits that reach out to much larger $R_\mathrm{max}$ and $z_\mathrm{max}$ than the high-[$\alpha$/Fe] stars. A similar conclusion was reached by \cite{Haywood2018InDR2}, who identified distinct ``wedges'' in this space. These wedges are also present when using the chemically defined halo sample of \cite{Chiba2000}, with updated astrometry from {\it Gaia} DR2, ruling out that they are due to a kinematic bias.  
\cite{Haywood2018InDR2} relate the wedges to the merger with Gaia-Enceladus and link them to impulsive heating of an ancient Milky Way disc. The authors suggest that the largest gap may hint at a phase transition in the assembly history of the Milky Way, from a significant to a quiescent accretion phase. A different view is presented by \cite{Amarante2020}, who argue that these wedges are due to transitions between orbital families. Based on smoothly sampled halo orbits, these authors attribute the structures in this space to ``a natural consequence of resonant effects''. 

In this work, we further explore the link between the $R_\mathrm{max} - z_\mathrm{max}$ space and orbital resonances, as mapped by frequency space. The space of orbital frequencies is particularly interesting for several reasons. Firstly, as shown by \citet{Gomez&Helmi2010} and \citet{Gomezetal2010}, individual streams associated with an accreted galaxy can be easily identified here. The time of accretion is quite precisely encoded in this space and can be determined from the separation between two adjacent streams in frequency space.
Secondly, \citet{Vallurietal2012} have argued that the strengths and locations of resonances for halo stars in frequency space are both dependent on the stellar distribution function as well as on the global shape of the halo. In principle, the orientation of the halo with respect to the disc can be determined from the distribution of stars in frequency space. Moreover, \citeauthor{Vallurietal2012} argue that the diffusion rates of the orbital frequencies can help in distinguishing between the true and an incorrect potential.
Finally, the mapping of frequency space is a powerful tool to establish the different orbital families present in an ensemble. For separable and non-degenerate potentials the frequencies are directly related to the actions. Frequencies have the advantage that they can be easily and precisely calculated numerically. Also in near-integrable potentials, for which the actions do not exist, the frequencies trace the different orbital families accurately \citep[e.g.][]{Laskar1993}.

In this work, we investigate the distribution of stars in the Solar vicinity in spaces associated with different orbital parameters - such as pericentre, apocentre, eccentricity, $z_{\rm max}$ and orbital frequencies - and explore the kinematics of the substructures identified in those subspaces. The paper is organised as follows. In Sec.~\ref{sec:data} we introduce the data and present the selection criteria applied. Then, in Sec.~\ref{sec:analysis}, we identify substructure in subspaces of different orbital parameters. In Sec.~\ref{sec:freq_analysis} we perform an orbital frequency analysis and show how the structures in the various subspaces, in particular, $R_\mathrm{max} - z_\mathrm{max}$ are linked to resonances in frequency space. 
We reflect on these results in Sec.~\ref{sec:discussion}, and present our conclusions in Sec.~\ref{sec:conclusions}.

%% file: Data.tex
\section{Data}\label{sec:data}

We use the subset of {\it Gaia} DR2 with line-of-sight velocities \citep{GaiaDR2_disckinematis_Katzetal2018}. This subset, also referred to as the \textit{Gaia} RVS (Radial-Velocity Spectrometer) subset, of more than $7$ million stars thus contains accurate measurements of the 3D positions and 3D velocities of the stars. \textit{Gaia} DR2 parallaxes are known to suffer from systematic parallax errors that vary with position, G-band magnitude, and stellar type. 
Therefore we use Bayesian distances $\hat{d}$ derived by \citet{McMillan2018_simpleGaiaDR2distances_RVSsubset}. The author has derived distances by assuming an overall parallax offset of $-29~{\rm mas}$ with an RMS error of $43~{\rm mas}$ for the \textit{Gaia} RVS subset as found for the full \textit{Gaia} DR2 release \citep{GaiaDR2_astrometricsolution_Lindegren2018}. We focus here on a local sample with $d<2.5$~kpc of high-quality parallaxes (${\tt parallax\_over\_error}>5$). These quality criteria leave us with a sample of $5~015~006$ stars. 

For these stars, we compute positions and velocities in a Galactocentric cylindrical coordinate system ($R$, $z$, $\phi$, $v_R$, $v_z$, $v_\phi$). We place the solar position at $R_\odot = 8.2$~\text{kpc} \citep{McMillan2017}\footnote{This value is consistent with \citet{Gravityetal2019_Rsun} who have recently determined $R_\odot = 8178\pm13_\mathrm{stat.}\pm22_\mathrm{sys.}$ pc.} and $z_\odot = 0.014$~\text{kpc} \citep[][]{Binneyetal1997}. 
The Local Standard of Rest (LSR) velocity, is for consistency also taken from \citet[][]{McMillan2017} (i.e. $v_\mathrm{c}(R_\odot) = 232.8$~\text{km/s}). We further set the motion of the Sun relative to the LSR to $(U,V,W)_\odot = (11.1, 12.24, 7.25)$~km/s \citep{Schonrichetal2010}. Here $U$ expresses radially inward motion, $V$ into the direction of Galactic rotation, and $W$ perpendicular to the Galactic plane and towards the Galactic north pole. We further define $v_\phi$ such that it is positive in the sense of Galactic rotation. 

\begin{figure}[t] \centering 
\includegraphics[width=\hsize]{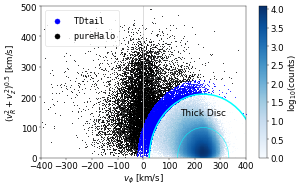} 
\caption{Velocity distribution of the stars in our sample, highlighting the various kinematic selections. A logarithmic map of star counts is shown in the background (the smooth blue map). Stars within the cyan semi-circles are considered to belong to the thick disc (TD). The stars outside the outermost cyan curve are considered halo stars, they are split into a pure halo and a thick disc tail sample.} 
\label{fig:Nstars_2d} 
\end{figure}

In this work, we focus on nearby thick disc (TD) and halo samples. 
Stars belonging to these samples are ``isolated'' based on their kinematics \citep[also see][]{Vennetal2004, Bensbyetal2014, Postietal2018}. We select stars that satisfy {$v_\mathrm{min} < |\vec{v}(\vec{x}) - \vec{v_\mathrm{c}}(R_\odot)| < v_\mathrm{max}$}. For the thick disc sample, we choose $v_\mathrm{min}^\mathrm{thick} = 100$~km/s and $v_\mathrm{max}^\mathrm{thick} = 210$~km/s, while for the halo we only set a minimum speed ($v_\mathrm{min}^\mathrm{halo} = 210$~km/s). These selections are illustrated in Fig.~\ref{fig:Nstars_2d}. The thick disc sample comprises $216~672$ stars and the halo sample $17~704$ stars. For this traditional definition of the halo, this consists of two main components, namely a hot thick disc and a proper halo \citep{GaiaDR2_HRdiagram_Babusiauxetal2018, Koppelmanetal2018_blob, Haywood2018InDR2, DiMatteo2018, Gallartetal2019}. 
Therefore, we split the halo sample further as shown in Fig.~\ref{fig:Nstars_2d} into a ``thick disc tail" ({\tt TDtail}, blue dots) and a ``pure halo" sample ({\tt pureHalo}, black dots). We set the boundary at $v_\mathrm{min}^\mathtt{pureHalo} = 260$~km/s, such that the {\tt pureHalo} sample contains $10~370$ stars and the {\tt TDtail} sample $7~334$ stars.

\begin{figure*}[ht!] \centering
	\includegraphics[width=0.99\textwidth]{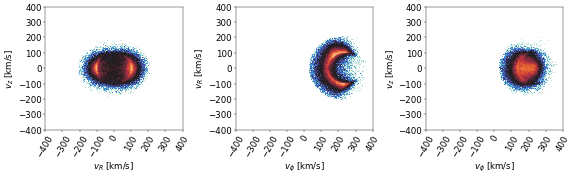}\\%
	\includegraphics[width=0.99\textwidth]{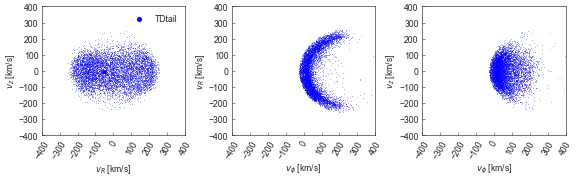}\\%
	\includegraphics[width=0.99\textwidth]{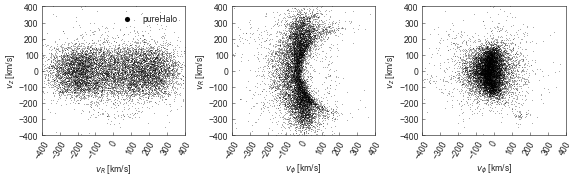}\\%
	\caption{Different projections of velocity space for the various subsets considered in this chapter: thick disc, thick disc tail, and pure halo from top to bottom. The lower density of stars present in these top panels for small values of $v_R$ and $v_z$ is due to the exclusion of thin disc stars in our selection. Note that even our {\tt pureHalo} subset contains some stars with ``hot'' thick disc-like kinematics (as evidenced by the arch-like structure in the bottom central panel).}
\label{fig:vel_vel_difpopulations} 
\end{figure*}

%% file: Results.tex
\section{Results: phase-space and integrals of motion}\label{sec:analysis}

For most of the analysis presented in this paper, we adopt the Milky Way (MW) potential determined by \citet{McMillan2017}, unless stated otherwise. This axisymmetric potential consists of a stellar thin and thick disc, H$_{\rm I}$ gas disc, molecular gas disc, a flattened bulge and a spherical halo component (NFW). This system has a virial mass of 
$1.3 \times 10^{12} M_{\odot}$, with $5.43 \times 10^{10} M_{\odot}$ in stars. 

We integrate the orbits of halo stars in this potential for $80$~Gyr, and $40$~Gyr for the thick disc stars. These long timescales are chosen because of our interest in computing the orbital frequencies, see Sec.~\ref{sec:freq_analysis}. Estimating the frequencies numerically requires densely sampled orbits with many periods, typically more than 20. Halo stars are integrated for a longer time than thick disc stars because the latter in general have much shorter orbital periods.
The orbits are integrated using \texttt{AGAMA} \citep{Vasiliev2019_AGAMA}, which uses an adaptive time-step (it uses the 8th order Runge-Kutta {\tt DOP853} integrator). We check that the energies of the stars do not deviate more than $0.01\%$ of their initial values. The data is output at fixed time intervals of roughly $1.25$~Myr and $2.5$~Myr for the halo and thick disc samples, respectively.


\subsection{Velocity space}\label{sec:velocityspace}

The local stellar distribution is enriched by several moving groups that are apparent as clumps in velocity space ($v_R$, $v_z$, $v_\phi$). These structures might be related to stars born together or may be of dynamical nature \citep[e.g. see][on the Outer Lindblad Resonance (OLR) mode due to the Galactic bar]{Antojaetal2018}. 

Figure~\ref{fig:vel_vel_difpopulations} shows the velocity distribution of the three subsets that we analyse, as indicated by the insets. As clearly apparent from this figure, 
these distributions are affected by the kinematic selection criteria used. Note also how the top (thick disc, TD) and middle row ({\tt TDtail}) look similar, whereas the bottom row ({\tt pureHalo}) shows a different distribution, mostly in the middle and right columns corresponding to $v_\phi$ vs.~$v_R$ and $v_\phi$ vs.~$v_z$, respectively. The {\tt TDtail} component (middle row) consists of stars with ``hotter'' thick disc-like kinematics, and it is likely the result of the proto-disc being puffed up during the merger with Gaia-Enceladus \citep{Helmietal2018, DiMatteo2018}. 

The strong radially elongated feature at $v_\phi = 0$ in the {\tt pureHalo} sample (middle panel) is indicative of the debris of Gaia-Enceladus \citep[c.f. ``the sausage'',][]{Belokurovetal2018}. The top and bottom parts of the crescent shape at $v_\phi \approx 100$~km/s imply that this sample still contains a small fraction of heated thick disc stars. It can be seen as an extension of the distribution shown in the middle panel of the {\tt TDtail} (middle row).

\begin{figure*}[ht!] \centering
	\includegraphics[width=0.9\textwidth]{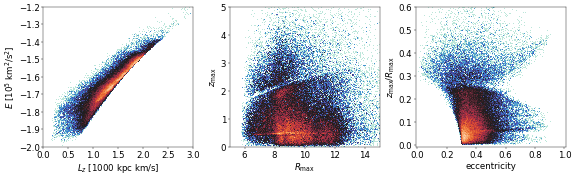}\\%
	\includegraphics[width=0.9\textwidth]{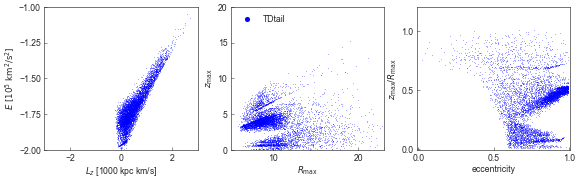}\\%
	\includegraphics[width=0.9\textwidth]{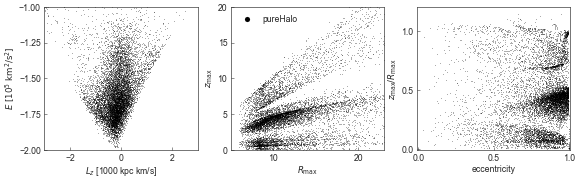}\\%
	\caption{Orbital parameters spaces for the different subsets defined in Sec.~\ref{sec:data}; from top to bottom: thick disc, thick disc tail, and pure halo. The left panels correspond to the most often used space to identify merger debris (i.e. $L_z,\,E$). The substructure seen near $L_z=250$~kpc km/s for highly bound energies in the middle-left panel is the globular cluster M4 (also known as NGC 6221), whose extent is largely the result of limitations in the {\it Gaia} DR2 astrometry for very dense stellar fields. The subspaces shown in the middle and right panels depict a large amount of structure, which appears to be independent of the subset or population considered. Note that the top row shows a smaller dynamical range than the other two rows.}
\label{fig:orbitalparameters_difpopulations} 
\end{figure*}%
 In the bottom, left panel we also see a clear overdensity of stars in an arch near $v_z \sim \pm 250$~km/s, for a range of values of $v_R$. These stars define the two clumps in the bottom right panel seen at  $(v_\phi \sim 120~{\rm km/s},v_z \sim \pm 250~{\rm km/s})$, they correspond to the Helmi streams \citep{Helmietal1999_HelmiStreams}.


\subsection{Characteristic orbital parameters}\label{sec:integralsofmotion}

In a static axisymmetric potential, such as the one considered here, the integrals of motion energy ($E$) and angular momentum in the $z$-direction ($L_z$) are conserved. Although stars that share a common origin phase-mix as they evolve in the Galactic potential \citep{Helmi&White1999}, and this results in complex configurations in phase-space \citep{Tremaine1999}, their distributions in, for example, $E$-$L_z$ space remain clumped nonetheless.

We show the distribution of the stars in the three samples in $E-L_z$ space in Fig.~\ref{fig:orbitalparameters_difpopulations}, see the left column. The middle and right column show a combination of other orbital parameters that are sometimes used to identify substructure. The {\tt pureHalo} sample (bottom row) shows, by far, the largest spread in the orbital parameters (note in particular the larger dynamical range w.r.t. the top row). Its distribution is dominated by stars near $L_z \sim 0$~kpc km/s, which is the debris identified as Gaia-Enceladus \citep{Koppelmanetal2018_blob, Helmietal2018}. Besides this prominent structure, we see contamination from the hot thick disc and several small overdensities such as that tentatively linked to an accretion event labelled Sequoia ($L_z \sim -2000$~kpc km/s, $E \sim -120000$ km$^2$/s$^2$), and Thamnos ($L_z \sim -1200$~kpc km/s, $E \sim -170000$ km$^2$/s$^2$) \citep[see][for more details]{Myeongetal2019, Koppelmanetal2019}.

\begin{figure*}[ht!] \centering
	\includegraphics[width=0.99\textwidth]{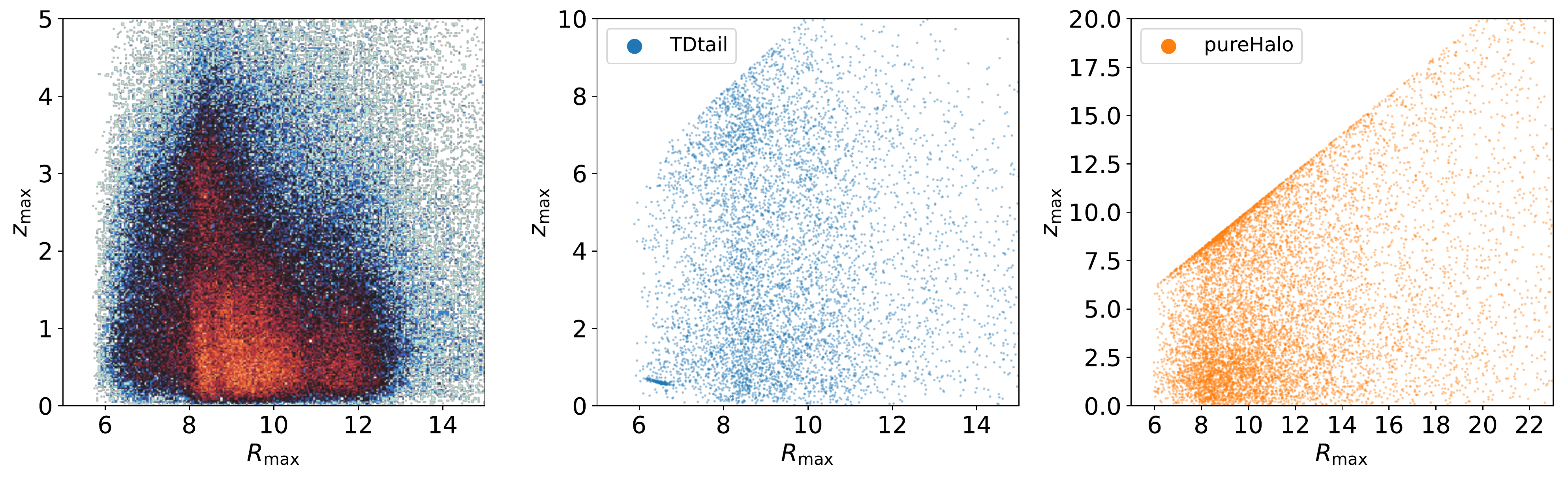}
    \caption{$R_\mathrm{max}$~vs~$z_\mathrm{max}$ for the thick disc (left), thick disc tail (middle), and pure halo (right) populations, computed for an integrable 2-component Galactic St\"{a}ckel potential. In all subsets the amount of structure has drastically decreased. The clear overdensity of stars in the thick disc tail sample and at $R_\mathrm{max}\sim6.5$~kpc and $z_\mathrm{max}\sim0.5$~kpc corresponds to the globular cluster M4 and is thus of no specific interest.}
\label{fig:stackel_orbitalparameters_difpopulations} 
\end{figure*}%

In middle panels of Fig.~\ref{fig:orbitalparameters_difpopulations} we plot 
$R_\mathrm{max}$~vs.~$z_\mathrm{max}$, while in the right panels we show the eccentricity vs. $z_\mathrm{max}/R_\mathrm{max}$ (i.e. a proxy for orbital inclination). The eccentricity is defined as $e=[r_a - r_p] / [r_a + r_p]$, where $r_p$ and $r_a$ denote the orbital pericentre and apocentre (measured as the minimum and maximum galactocentric distances over the whole integration interval). 
Particularly in these orbital parameters spaces, we see a large amount of structure. These structures look similar to the ``wedges'' reported by \citet{Haywood2018InDR2} and computed for a different Milky Way potential than used here. Figure~\ref{fig:orbitalparameters_difpopulations} shows that the structures are present in all the star samples explored, although with different prominence. The fact that they are present independently of the population indicates that the phenomenon that causes them must be of global nature and is unlikely to be due to (ancient) accretion events. 

The shape and exact location of the structures shift slightly when changing the Galactic potential, but they appear to be generic. For example, we have also integrated our samples in the potentials provided by \cite{Piffletal2014_constraininghalowithRAVE} and the \texttt{MWPotential2014} \citep{Bovyetal2015_Galpy} and found similar features \citep[see also][]{Haywood2018InDR2}. 

To gain further insight we also decided to explore a less generic but integrable potential, namely a Milky Way-like potential of St\"{a}ckel form \citep[derived from][]{Batsleer&Dejonghe1994, Famaey&Dejonghe2003}. Potentials of this form are separable and contain only regular orbits. Figure~\ref{fig:stackel_orbitalparameters_difpopulations} shows the distributions of the subspace $R_\mathrm{max}$ vs.~$z_\mathrm{max}$ for the stars in our various subsets now determined using a two-component St\"{a}ckel Milky Way potential of \citet{Famaey&Dejonghe2003}. This potential has a total mass of $M_{\rm tot} = 4\times 10^{11}~{\rm M_\odot}$, a disc mass with a fraction of $k=0.11$ of $M_{\rm tot}$, scale length $r_a = 7~{\rm kpc}$, and a flattening of the disc and halo of $\epsilon_d = 75$ and $\epsilon_h = 1.02$, and provides a good representation of for example, the Galactic rotation curve. Fig.~\ref{fig:stackel_orbitalparameters_difpopulations} reveals rather smooth and featureless distributions, especially in comparison to the previous figure. This implies that the structures seen in Fig.~\ref{fig:orbitalparameters_difpopulations} must be induced by the non-integrability of the generic Galactic potentials used. The structures could thus be due to resonances and the result of chaotic diffusion, a hypothesis that we explore further in Sec.~\ref{sec:freq_analysis}. 

\subsection{Comparison to known substructures}\label{sec:substructures}
To further strengthen this preliminary conclusion 
we investigate the distribution of stars associated with known accreted substructures in the orbital parameters spaces just explored. Many candidate substructures have been identified since \textit{Gaia} DR2, such as Gaia-Enceladus \citep{Helmietal2018}, Sequoia \citep{Myeongetal2019}, and Thamnos \citep{Koppelmanetal2019}, and also confirmed such as the Helmi Streams \citep{Helmietal1999_HelmiStreams, Koppelmanetal2018_Helmistreams}. 

\begin{figure*}[ht!] \centering
	\includegraphics[width=0.99\textwidth]{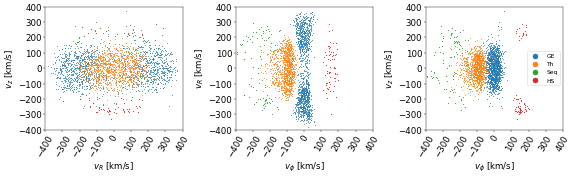}\\
	\includegraphics[width=0.99\textwidth]{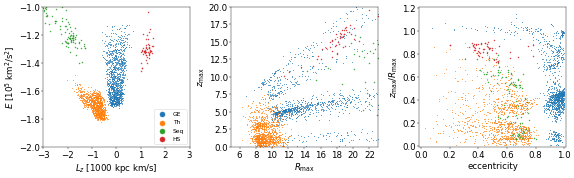}\\%
	\caption{Distribution of accreted substructures in the {\tt pureHalo} subset following the identification by \citet{Koppelmanetal2019}, in velocity space (top) and in orbital parameters spaces (bottom).}
\label{fig:substructure_halo} 
\end{figure*}%

We take the stars that belong to these substructures according to \citet{Koppelmanetal2019} and take their intersection with our {\tt pureHalo} sample. In this way we find 
$1186$ Gaia-Enceladus stars, $54$ members of the Helmi Streams, $92$ in Sequoia, and $1098$ Thamnos stars. 

The velocity distribution of the stars in these various substructures is shown in the top panels of Fig.~\ref{fig:substructure_halo}. The bottom left panel directly reflects the selection criteria used to identify them (where a lower limit in the energy $E$ has been somewhat artificially imposed on the Gaia-Enceladus stars to reduce overlap with the bulge and the tail of the thick disc \citep[e.g.][]{Feuillet2020}. This constraint depletes the number of stars with small $v_R$ and eccentricities $e < 0.8$). On the other hand, the middle and right-hand side panels of the bottom row show that these accreted substructures are also themselves split into substructures or wedges, just like the stars in the various subsets considered in the previous section. For example, the nearly horizontally aligned gaps in ($R_\mathrm{max}, z_\mathrm{max}$)-space continue to be very prominent. Given the (tentative) range of accretion times for the various objects ($\sim$6 to $10$~Gyr ago), and that their debris seems to be influenced in the same way, this analysis further supports the idea that the origin of the features seen in Fig.~\ref{fig:orbitalparameters_difpopulations} must lie in a global phenomenon affecting the whole Galaxy.

%% file: Analysis.tex
\section{Analysis: Orbital Frequencies}\label{sec:freq_analysis}

\subsection{Methods}\label{Methods}

In this section, we perform a frequency analysis to explore whether resonances might play a role in the substructures seen in the orbital parameters spaces. We will numerically estimate the frequencies using the \texttt{SuperFreq} package\footnote{\url{https://superfreq.readthedocs.io/en/latest/}} \citep[][see also \citealt{Price-Whelanetal2016}]{Price-Whelan2015SuperFreq:Orbit}, which is a Python implementation very similar to the Numerical Analysis of Fundamental Frequencies (or NAFF) code written by \citet{Valluri&Merritt1998,Vallurietal2010, Vallurietal2012}, which in itself is an implementation of the NAFF technique of \citet[][see also \citealt{Wangetal2016}, who provide a comparison between the NAFF algorithm and another commonly used orbital frequency and classification code by \citealt{Carpintero&Aguilar1998}]{Laskar1990, Laskar1993}. The software finds the fundamental frequencies by computing the Fourier spectra for the phase space coordinates (more precise: of a complex time-series) used to describe the orbit. Since the orbits are computed in an axisymmetric potential we follow \citet{Vallurietal2012} by using a slightly different form of the cylindrical polar coordinates, namely the Poincar\'{e}'s symplectic polar variables. In this case, the complex time-series are chosen as: $f_R = R + i v_R$, $f_z = z + i v_z$, $f_\phi = \sqrt{2L_z}\left[\cos(\phi) + i \sin(\phi)\right]$. The frequencies in $R$ and $z$ are defined to be positive, while the sign of the $\phi$-frequency corresponds to the direction of rotation (i.e. positive when aligned with the Galactic rotation).\footnote{By definition, the fundamental frequencies are non-zero. However, a zero frequency line in the spectrum of a coordinate can indicate that the orbit's centre is not on the origin. For example, a zero-frequency in $z$ indicates an asymmetric orbit such as a ``banana'' orbit.}

In \texttt{SuperFreq}, the first fundamental frequency is the non-zero frequency with the highest amplitude. The algorithm then continues by moving down along the remaining frequency lines in the Fourier spectrum corresponding to the other coordinates, in order of decreasing amplitude. The next fundamental frequency must be different from the first one found, and so on. These fundamental frequencies $\vec{\Omega}=(\Omega_R,\Omega_z,\Omega_\phi)$ are not necessarily equal to the dominant frequencies $\vec{\omega}$ in the Fourier spectra of each coordinate. The fundamental frequencies found are also not necessarily equal to the frequencies in which the angle variables in action-angle coordinates vary with time. For regular orbits, the recovered fundamental frequencies will, however, be a linear combination of these ``more fundamental'' (action-angle) frequencies \citep{Vallurietal2012}.

Resonant orbits are those for which the fundamental frequencies are commensurable. 
That is, a resonance is defined to exist if $\vec{n} \cdot \vec{\Omega} = 0$, for any vector $\vec{n}=(n_z, n_R, n_\phi)$ comprising integer numbers (where at most one element is zero). In theory there are an infinite number of resonances for combinations of arbitrarily large vector elements $n_i$, here we only consider those with $|n_i|\leq5$.
In a spherical potential, the true ``fundamental'' frequencies corresponding to the nature of the potential satisfy: $\Omega_r/2 \leq |\Omega_\phi| \leq \Omega_r$ \citep{Binney&Tremaine2008}, where the lower limit corresponds to any orbit in a homogeneous sphere (harmonic oscillator), and the upper limit corresponds to orbits in a Kepler potential (point mass). In the epicyclic approximation, $\Omega_\phi/\Omega_r = 1/\sqrt{2} \simeq 0.707$ for a flat rotation curve [$v_\textrm{circ}(r) = \textrm{cst}$]. In an axisymmetric disc potential, vertical oscillations for most disc stars are expected to have the shortest periods ($\Omega_z$ is the largest in $\vec{\Omega}$). 

The NAFF methodology provides reliable frequencies for regular orbits which have been traced for $20-30$ periods, so we limit our analysis to stars with a minimum of $20$ orbital periods. All thick disc stars and almost all halo stars (99.76\%) are integrated long enough to satisfy this criterion. Most halo stars are integrated for $100-600$ orbital periods, while most thick disc stars for $150-300$ orbital periods.

\begin{figure*}[ht!] \centering
	\includegraphics[width=0.99\textwidth]{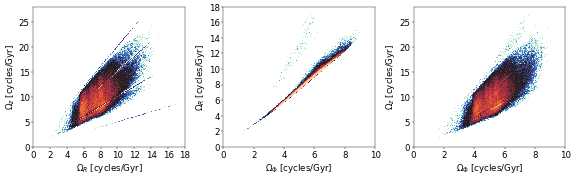}\\%
	\includegraphics[width=0.99\textwidth]{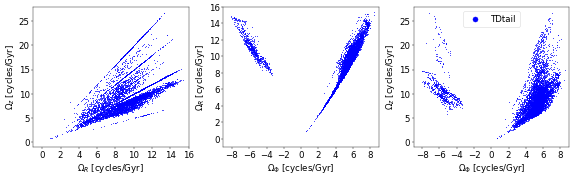}\\%
	\includegraphics[width=0.99\textwidth]{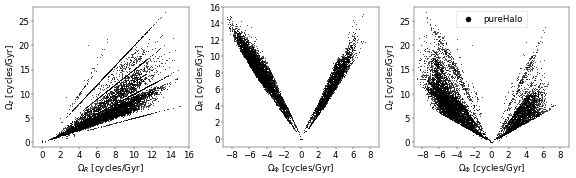}\\%
	\caption{Frequency maps for the different populations. From top to bottom: thick disc, thick disc tail, and pure halo population. From left to right: cycling through the combinations of frequencies. Notice the presence of resonances (especially in the left panels) in the form of straight lines, and of depleted regions around them.}
\label{fig:freq_freq_difpopulations} 
\end{figure*}

\begin{figure*} 
\centering
	\includegraphics[width=0.65\textwidth]{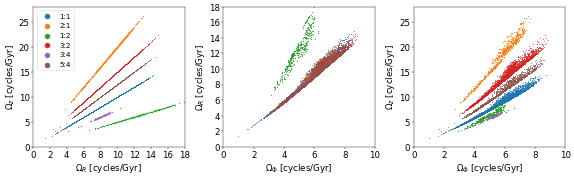}\\%
	\includegraphics[width=0.65\textwidth]{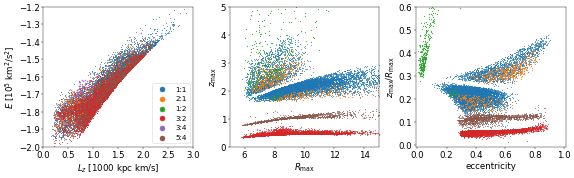}\\
	\includegraphics[width=0.65\textwidth]{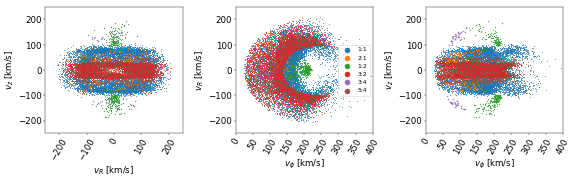}\\%
    \caption{Selected resonances in the thick disc subset and their mapping onto other informative subspaces. Top row: Frequency maps highlighting the main resonant families. Middle row: Characteristic orbital parameters colour-coded according to the different resonances. Bottom row: Their distribution in velocity space.}
    \label{fig:freq_freq_resonance_thick} 
\end{figure*}

\subsection{Results}\label{sec:freq_analysis_results}
We commence the frequency analysis by showing in Fig.~\ref{fig:freq_freq_difpopulations}, the distribution of fundamental orbital frequencies for each of the populations/subsets considered. The top row shows a density map because of the size of the sample, the colours indicate the logarithm of the number of stars with bright orange being the densest and blue being the least dense.
Resonances are clearly discernible in this space as well-populated, thin straight lines. Most of the resonances are present in the $\Omega_R$~vs.~$\Omega_z$ space (left column). The regions around these resonances are depleted of stars, which is common for non-integrable potentials and indicates the presence of chaotic zones \citep[e.g. see Fig.~1 of][for an illustration of this phenomenon]{Price-Whelanetal2016}. There are no very obvious resonances in $\Omega_\phi$~vs.~$\Omega_R$ \citep[middle panels, although these might be expected for thin disc stars, particularly in relation to the Galactic bar, see][]{Dehnen2000}. 

We find that roughly 10\% of the stars in the thick disc sample are on resonant orbits in $\Omega_z/\Omega_R$. For the {\tt TDtail} and for the {\tt pureHalo} samples, these fractions are much higher, namely {25\%} and {28\%} respectively. Stars are said to belong to a certain resonance if the orbital frequency ratio for the pair of coordinates deviates at most $0.001$ from the rational number of the resonance under consideration. To establish if the resonances are related to the structures we identified in the previous section we will now study the most dominant resonances in each of the subsets. 

A selection of the most dominant resonances in the TD sample is shown in Fig.~\ref{fig:freq_freq_resonance_thick} (top row). These are the $\Omega_z:\Omega_R=$~1:1 (blue), 2:1 (yellow), 1:2 (green), 3:2 (red), 3:4 (purple), 5:4 (brown) resonances. The percentage of stars in each of these resonances are listed in
Table~\ref{tab:res}. 
The middle row of Fig.~\ref{fig:freq_freq_resonance_thick} shows that the stars on these resonances occupy specific regions in the spaces of orbital parameters. There is a strong connection between the frequencies and features seen in these spaces. The overdensities first shown in Fig.~\ref{fig:orbitalparameters_difpopulations} correspond directly to some of the highlighted resonances. On the other hand, depleted regions and gaps in the orbital parameters spaces indicate regions of unstable orbits. Such orbits are likely trapped by the resonances explored here, leaving empty regions around the corresponding frequencies. This is for example clearly seen for the $\Omega_z:\Omega_R=$~1:2 (in green) and  $\Omega_z:\Omega_R=$~3:2 (in red) resonances.
The bottom panels of Fig.~\ref{fig:freq_freq_resonance_thick} show that stars on a given resonance have a broad range of velocities. However, some resonances (e.g. the 1:2) occupy only a small region in certain projections of velocity space.

\begin{table}
\caption{Percentage of stars in the various subsamples associated with the listed $\Omega_z:\Omega_R$ resonances. The percentages are calculated with respect to the entire sample - not just relative to the stars on resonances.}
\begin{tabular}{crrrrrrr}
\hline
sample &  1:1 & 2:1 & 1:2 & 3:2 & 3:4 & 5:4 & 5:2 \\
\hline
TD [\%] & 5 & 0.3 & 0.2 & 3.6 & 0.03 & 0.7 & 0 \\
TDtail [\%] & 10.9 & 4.2 & 0.6 & 4.8 & 3.1 & 1.9 & 0 \\
pureHalo [\%] &  10.1 & 1.7 & 6.2 & 3.9 & 4.3 & 1.8 & 0.4 \\ 
\end{tabular}
\label{tab:res}
\end{table}

Figure~\ref{fig:freq_freq_resonance_thick} shows that most of the resonances can be associated with a contiguous region in the orbital parameters spaces. However, this figure also reveals a higher complexity for the $\Omega_z:\Omega_R=$~1:1 and the $\Omega_z:\Omega_R=$~2:1 resonances. The stars on these families are found in two regions that surround the most prominent gap of Fig.~\ref{fig:orbitalparameters_difpopulations}, located near $R_\mathrm{max}\sim8$~kpc and $z_\mathrm{max}\sim2$~kpc. This potentially means that these resonances are not solely responsible for this depleted region. We return to this point in the next section.

\begin{figure*}\centering
	\includegraphics[width=0.65\textwidth]{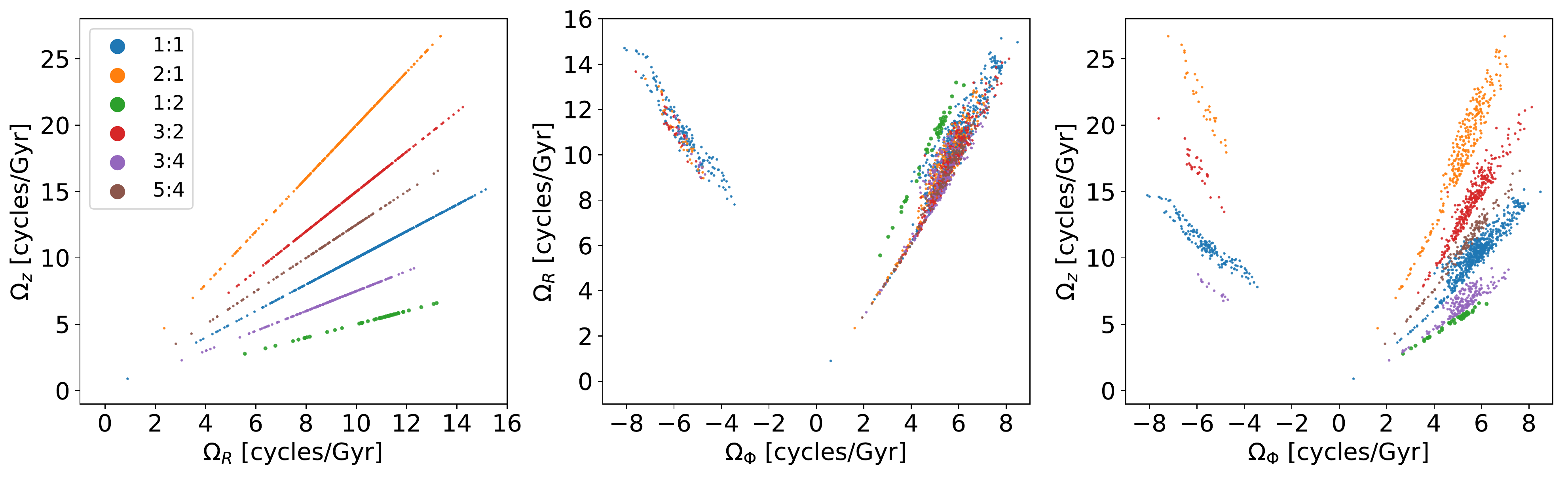}\\ 
	\includegraphics[width=0.65\textwidth]{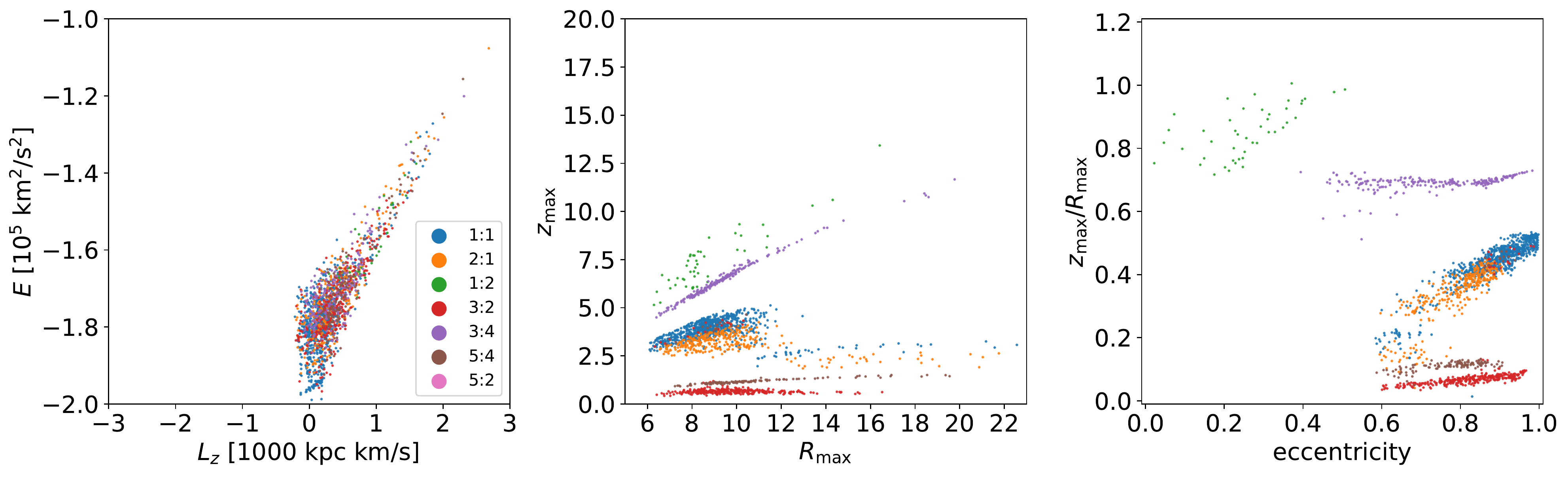}\\
	\includegraphics[width=0.65\textwidth]{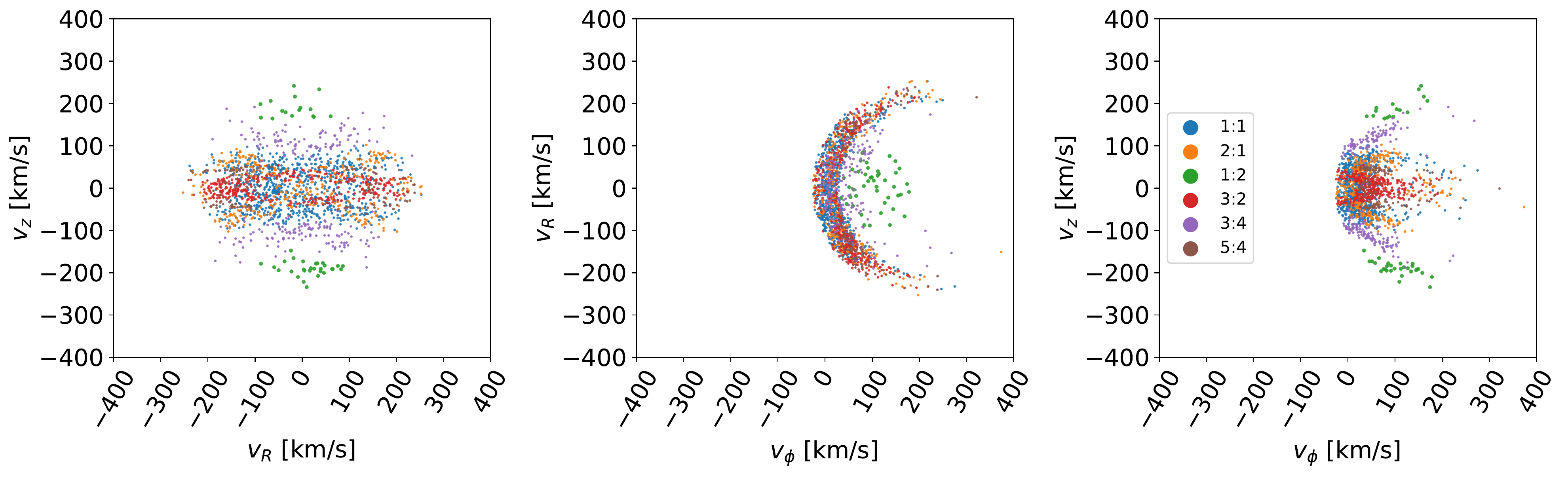}\\%
	\caption{Similar to Fig.~\ref{fig:freq_freq_resonance_thick} but now for the {\tt TDtail} sample.}
\label{fig:freq_freq_resonance_TDtail} 
\end{figure*}%
\begin{figure*}
\centering
	\includegraphics[width=0.65\textwidth]{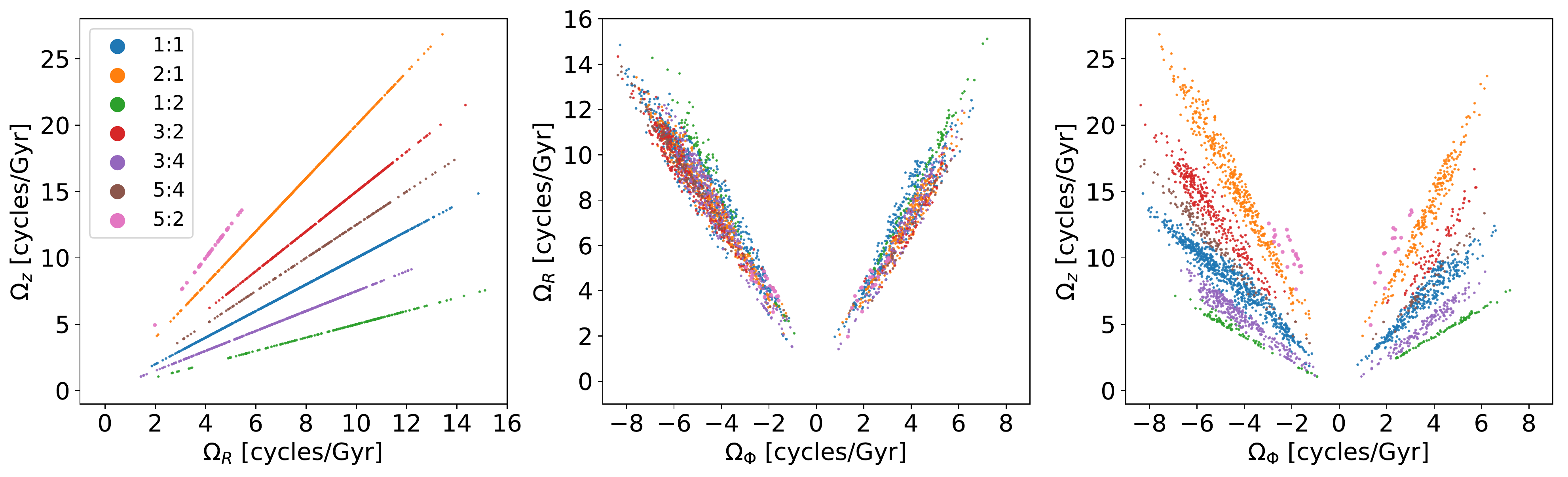}\\ 
	\includegraphics[width=0.65\textwidth]{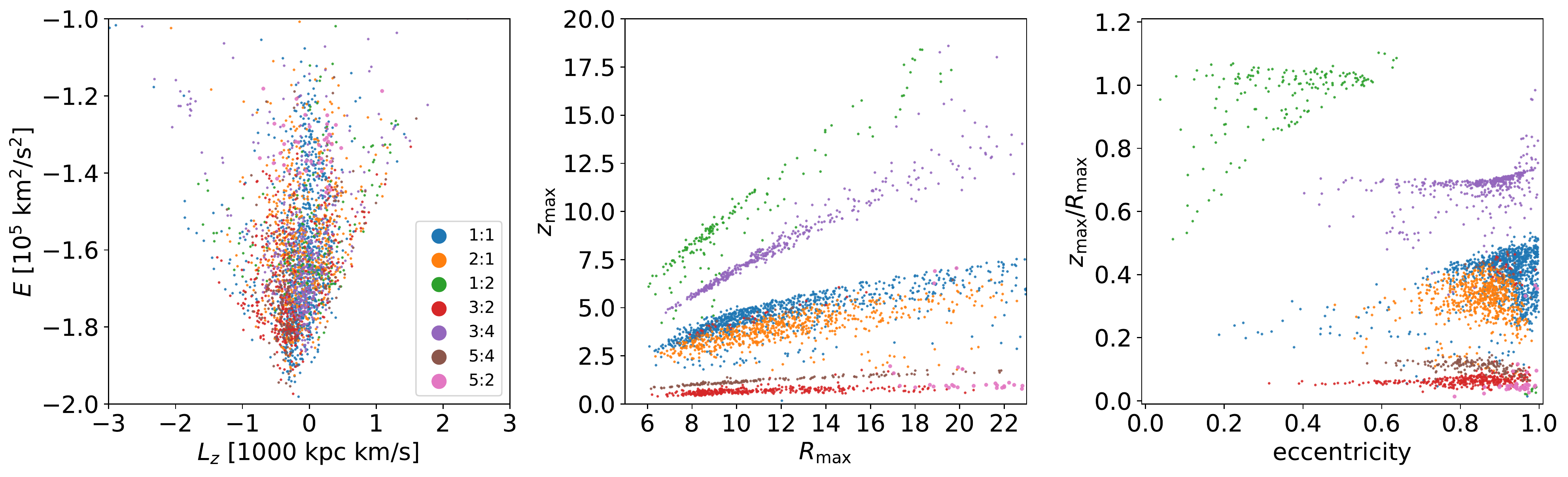}\\
	\includegraphics[width=0.65\textwidth]{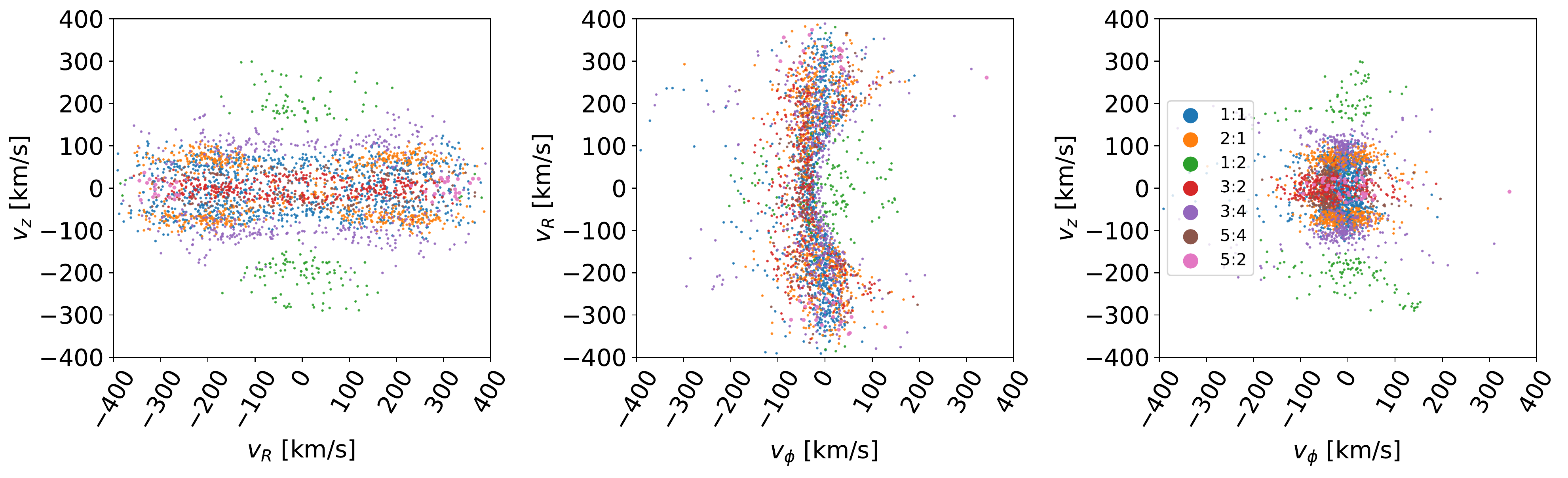}\\%
	\caption{Similar to Figs.~\ref{fig:freq_freq_resonance_thick} and~\ref{fig:freq_freq_resonance_TDtail} but now for the {\tt pureHalo} sample.}
\label{fig:freq_freq_resonance_pureHalo} 
\end{figure*}%

Following the same procedure for the halo samples, we show separately the {\tt TDtail} and the {\tt pureHalo} samples in Fig.~\ref{fig:freq_freq_resonance_TDtail} and~\ref{fig:freq_freq_resonance_pureHalo} respectively. In addition to the 
$\Omega_z:\Omega_R$ resonances identified in the thick disc sample, we also find that the 3:4 and 5:2 being populated. Just like we have seen for the thick disc sample, the stars associated with the resonances also occupy specific regions in the orbital parameters spaces.

\subsection{Commonalities between subsamples}\label{sec:Commonalities}
 
In all three subsamples, we find similar families of resonances being dominant. The $\Omega_z:\Omega_R=$~1:1 is the most prominent resonance in all samples. In the thick disc sample, the other dominant resonance is the 3:2 resonance. For the {\tt TDtail} sample, this resonance is also the second most dominant frequency, but the 2:1 and 3:4 resonances are also prominent. In the {\tt pureHalo} sample, the second most dominant resonance is the 2:1, then followed by the 3:4 and 3:2 resonances.

\begin{figure*}[ht!] \centering
	\includegraphics[width=0.99\textwidth]{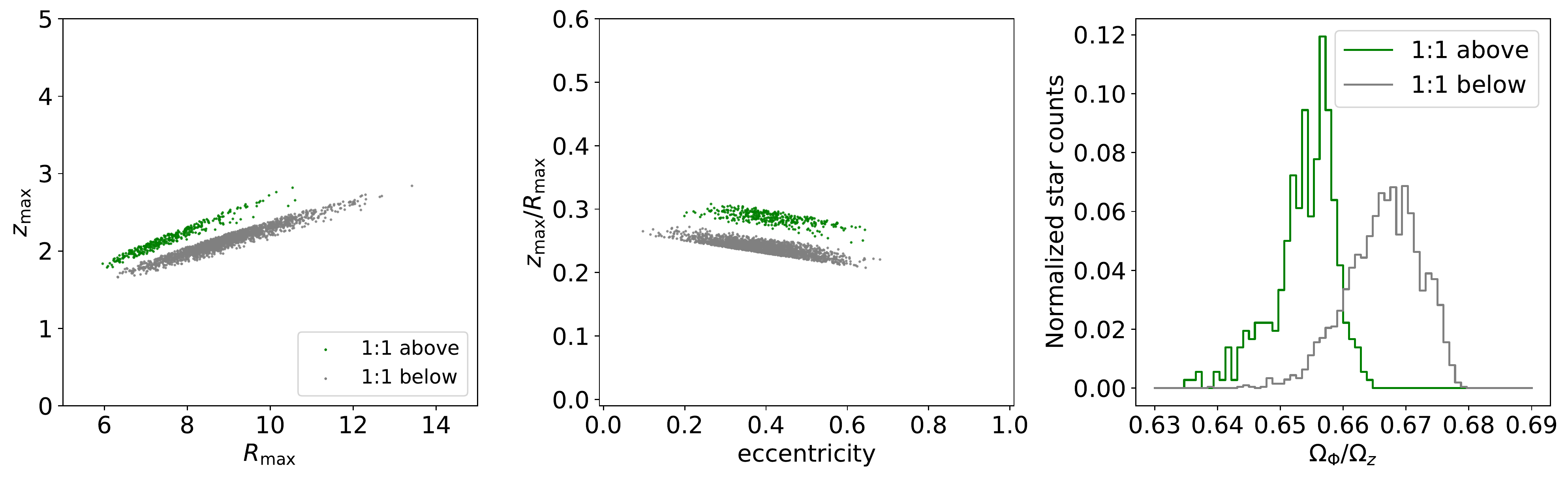} \\ 
    \caption{Selection of stars (left panel) around the gap in $R_\mathrm{max}$~vs.~$z_\mathrm{max}$ for the $\Omega_z:\Omega_R=$~1:1 resonance for stars in the thick disc sample. Their mapping to eccentricity vs.~$z_\mathrm{max}/R_\mathrm{max}$ and the distributions of their frequency ratios $\Omega_\phi/\Omega_z$ are respectively shown in the middle and right panels. This last panel shows that a splitting/bifurcation appears to occur for $\Omega_\phi/\Omega_z = $~2:3.}
\label{fig:Rmax_zmax_gap_thick} 
\end{figure*}

\begin{figure*}
\centering
	\includegraphics[width=0.99\textwidth]{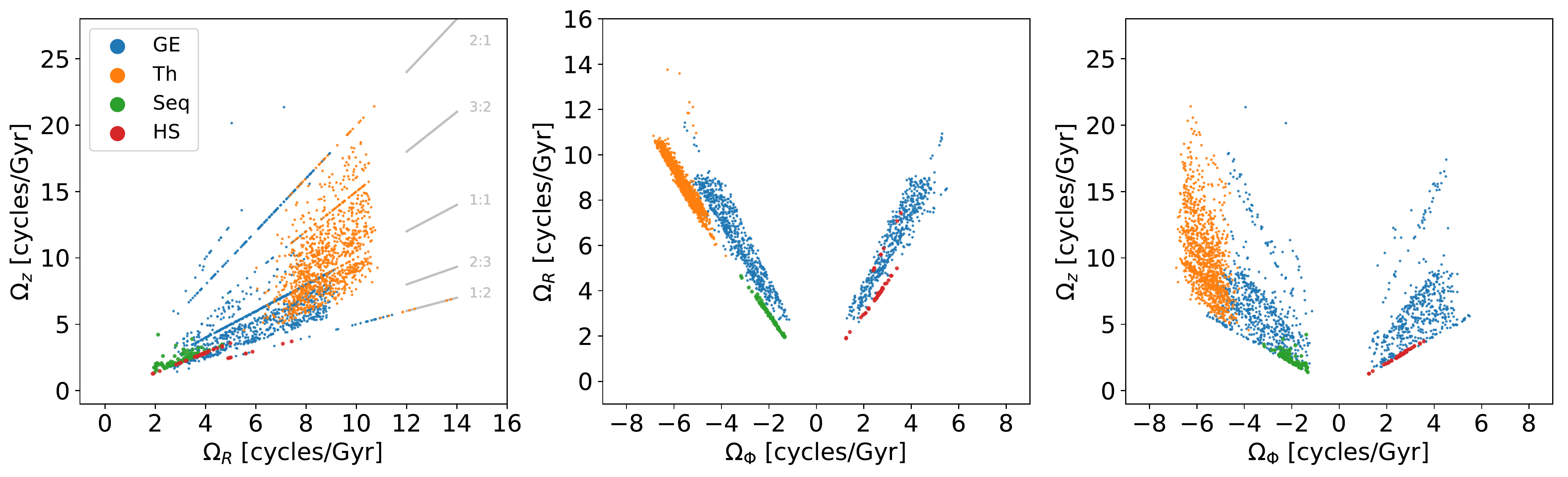}\\%
	\caption{Frequency maps of the known accreted substructures for the stars present in our {\tt pureHalo} sample. Some of the clearest resonances in the panel on the left have been marked with light gray lines.}
\label{fig:freq_freq_substructure_halo} 
\end{figure*}

We now return to the most prominent gap seen in Fig.~\ref{fig:orbitalparameters_difpopulations} and present in all subsamples, that is the gap that goes through $(R_\mathrm{max}, z_\mathrm{max})\sim(8,2)$~kpc. We focus for practical purposes on the thick disc subset. We have previously noted from the middle row panels of Fig.~\ref{fig:freq_freq_resonance_thick} that the gap also divides the stars associated with the resonances $\Omega_z:\Omega_R=$~1:1, 2:1, and 1:2 in two separate zones. This implies that
these resonances on their own can not fully explain the presence of a gap in the $R_{\rm max}$ vs $z_{\rm max}$ space. 

To gain further insight, we select stars around the gap for the $\Omega_z:\Omega_R=$~1:1 resonance, as shown in the leftmost panel of Fig.~\ref{fig:Rmax_zmax_gap_thick}. The rightmost panel of this figure shows that the stars in this region cluster around $\Omega_\phi:\Omega_z =$~2:3, but that those above the gap have $\Omega_\phi/\Omega_z < 2/3$, while those below have values greater than 2/3. Similar results are found for the other resonant families, with the gap seen in the characteristic orbital parameters of stars associated with the $\Omega_z:\Omega_R=$~2:1 resonance, being related to $\Omega_\phi:\Omega_z =$~1:3. Overlapping chaotic zones, related to resonances, enhance chaotic diffusion \citep[e.g.][]{Laskar1993}. We surmise that the most prominent gap in $(R_\mathrm{max}, z_\mathrm{max})$ shown here is due to the 3D $(\Omega_\phi:\Omega_R:\Omega_z) = (2:2:3)$ resonance.

Figures \ref{fig:freq_freq_resonance_thick},
\ref{fig:freq_freq_resonance_TDtail}, and
\ref{fig:freq_freq_resonance_pureHalo} show that the velocity
distributions of stars associated with resonances reveal the presence
of clumps, which could potentially be confused with merger debris
(i.e. such as the Arcturus stream, see
\citealt{Navarroetal2004_Arcturus} or \citealt{Kushniruk&Bensby2019}).
A more recent example is the reported prograde stellar stream, Nyx,
with azimuthal velocities around $140$~km/s
\citep{Necib2019EvidenceVicinity}, which could also be due to the
Galactic bar acting on thick disc-like stars, but which we have not
considered here (e.g. see \citealt{Monarietal2013} or
\citealt{Antojaetal2015}). We note that, in most cases, the clumps
associated to resonances do not correspond to overdensities in
velocity space for the full data set, as can be seen from a comparison of 
 Figures \ref{fig:freq_freq_resonance_thick},  \ref{fig:freq_freq_resonance_TDtail}, and
\ref{fig:freq_freq_resonance_pureHalo} to Fig.~\ref{fig:vel_vel_difpopulations}.

\subsection{Known accreted substructures in frequency space}\label{sec:substructure in orbital parameter space}

We now explore the regions of frequency space occupied by the accreted
substructures introduced in Sec.~\ref{sec:substructures}. Their
kinematics and orbital parameters were shown in
Fig.~\ref{fig:substructure_halo}, while
Fig.~\ref{fig:freq_freq_substructure_halo} shows their distribution in
frequency space.

Fig.~\ref{fig:freq_freq_substructure_halo} reveals that Gaia-Enceladus
and Thamnos are seen to span rather large regions in frequency
space. They are affected by resonances just like the whole halo
subset. The Sequoia stars occupy an small region close to and around
the $\Omega_\phi:\Omega_R=$~-2:3 resonance. On the other hand, the Helmi
Streams (in red) occupy two relatively narrow regions in
$\Omega_\phi$~vs.~$\Omega_R$, and appear to be split also in
$\Omega_R$~vs.~$\Omega_z$, with one of the two groups being on the
$\Omega_z:\Omega_R=$~1:2 resonance. There are stars both from the
positive and negative $z$-velocity clumps on the 1:2 resonance, with
the associated stars being slightly colder kinematically. The
splitting into clumps in frequency space could potentially be due to
different wraps of the debris streams \citep{Gomez&Helmi2010}, but
clearly a much higher number of stars would be needed to confirm this
interpretation.

%% file: Discussion.tex
\section{Discussion}\label{sec:discussion}

The analysis carried out in this work shows that substructures present in spaces associated with the orbital parameters of stars
can be driven by properties of the gravitational potential in which the stars move. In particular, we have seen that a realistic (but non-integrable) Milky Way potential, with a disc and halo components, leads to the presence of well-populated orbital resonances. Trapping of orbits at resonances and chaotic zones lead to the depletion of stars in the regions around these resonances. 

These results do not strongly depend on the choice for the gravitational potential assumed for the Milky Way, provided this is generic enough (and not integrable, e.g. of St\"ackel form). 
We have found that also in two other commonly used Galactic potentials \citep{Piffletal2014_constraininghalowithRAVE, Bovyetal2015_Galpy} very similar substructures can be observed in $R_\mathrm{max}$~vs.~$z_\mathrm{max}$, and that they are associated with resonant families, and hence to the presence of non-integrability in the system.

Our analysis therefore directly shows that not all substructure is due
to accretion as often considered in the literature, nor to the
settling of the gravitational potential after major merger activity as
had been suggested by \citet{Haywood2018InDR2} \citep[as also
concluded by][]{Amarante2020}. Nonetheless, the characteristics of
substructures from merger events may sometimes be related to the
presence of resonances. For example, we have seen that some of the
stars in the Helmi streams appear to be close to a resonance,
$\Omega_z:\Omega_R\sim$~1:2. This perhaps explains why its stars are
distributed asymmetrically in velocity space (the stream with
$v_z < 0$ has more stars than that with $v_z > 0$). This asymmetry has
been used to constrain its time of accretion to approximately 5-8 Gyr
ago, which is a puzzling low value given that these stars are on
relatively bound orbits. Since stars near a resonance take longer to
spread out in space \citep{Vogelsbergeretal2008}, this can lead to an
underestimation of their time of accretion. This means that chaotic
dynamics may have to be considered when modelling the evolution of
tidal debris, as already hinted by \citet{Price-Whelanetal2015} in
their modelling of cold thin streams farther away in the halo.

Probably the best way to tell whether substructures are related to accretion events is via a chemical tagging analysis. Stars that originate from a satellite will follow characteristic tracks in chemical abundance space. On the other hand, stars that group together because of dynamical resonances have no reason to be chemically distinct from other stars. 

How stars populate different resonant families depends on their
distribution function and the gravitational potential in which they
move. In this work, we only considered that of the Milky Way, but a
recent analysis of nearby thin disc stars has shown that also the
Sagittarius dwarf plays a role in their dynamics
\citep[see][]{Antojaetal2018}. Interestingly, if we integrate the
orbit of Sagittarius' in the McMillan (2017) potential for 40 Gyr with
the initial conditions derived by
\citet{GaiaCollaboration2018Helmi}, and apply the \texttt{SuperFreq}
code to obtain its frequencies (i.e. we repeat the analysis done for
the stars in our sample), we find that Sagittarius falls on the
$\Omega_z:\Omega_R=$~1:2 and $\Omega_z:\Omega_\phi=$1:1 resonance. The
stars in our sample on the $\Omega_z:\Omega_R=$~1:2 resonance (green
points in
Figs.~\ref{fig:freq_freq_resonance_thick},~\ref{fig:freq_freq_resonance_TDtail},
and~\ref{fig:freq_freq_resonance_pureHalo}) define a rather distinct
branch in frequency space. They also have rather different velocities,
with high $v_z$ values. Although our integration did not include the
gravitational potential due to the Sagittarius dwarf, this
``coincidence'' suggests that its effect on the hotter components of
the Milky Way is non-negligible and would probably be even more
reinforced if we had considered it in our orbital
integrations. Furthermore, this is also the resonance where some of
the Helmi stream stars lie. Whether this has been induced by
Sagittarius, or indicates a link between the origin of the two
galaxies, is not clear but certainly deserve further investigations.

%% file: Conclusions.tex
\section{Conclusions}\label{sec:conclusions}

We have studied the dynamical properties of nearby stars in the thick disc and stellar halo using data from the 2nd data release of the {\it Gaia} mission. We have explored spaces of characteristic orbital parameters for a generic Galactic potential and found that these host a large amount of structure. Further analysis using orbital frequencies shows that these structures are due to the presence of resonant families \citep[as claimed also by][]{Amarante2020}, and to the depletion of orbits around them due to non-integrability. Therefore these structures reflect intrinsic properties of the gravitational potential of the Milky~Way. 

These findings are interesting on their own and highlight that a large number of stars in these components (nearly 30\% for the halo sample) are on resonances. This gives us hope to use them to pin-down more precisely the gravitational potential of the Milky Way. \citet{Vallurietal2012} argue that most stars will not be launched on regular tori and that therefore, chaotic diffusion should occur. The expectation would thus be that the fraction of stars on such irregular orbits will be the lowest for the true potential (as this would indicate self-consistency between the distribution function and the gravitational potential). 

Frequency analysis is also interesting because it can reveal the
individual streams originating in an accretion event
\citep{Gomez&Helmi2010}. Our analysis shows hints of such imprints,
but the number of stars is too small to reach meaningful
conclusions. Furthermore, it may be particularly challenging to identify
the individual streams or wraps as these may be disguised and trapped
by the resonances present in the phase-space of halo stars.